\documentclass[prl,twocolumn,superscriptaddress,showpacs]{revtex4-1}

\usepackage{amsmath}
\usepackage{amssymb}
\usepackage{graphicx}

\newcommand{\be}[1]{\begin{equation}\label{#1}}
\newcommand{\ee}{\end{equation}}
\newcommand{\ba}[1]{\begin{eqnarray}\label{#1}}
\newcommand{\ea}{\end{eqnarray}}

\usepackage[svgnames,table,hyperref]{xcolor}

\begin{document}

\title{Mode Conversion and Period Doubling 
at Plasma-$\beta$ Unity in an Alfv\'en-Wave Experiment with Liquid Rubidium}

\author{F. Stefani}
\email{F.Stefani@hzdr.de}
\affiliation{Helmholtz-Zentrum
Dresden-Rossendorf, Bautzner Landstr. 400, D-01314 Dresden, Germany}
\author{J. Forbriger}
\affiliation{Helmholtz-Zentrum
Dresden-Rossendorf, Bautzner Landstr. 400, D-01314 Dresden, Germany}
\author{Th. Gundrum}
\affiliation{Helmholtz-Zentrum
Dresden-Rossendorf, Bautzner Landstr. 400, D-01314 Dresden, Germany}
\author{T. Herrmannsd\"orfer}
\affiliation{Helmholtz-Zentrum
Dresden-Rossendorf, Bautzner Landstr. 400, D-01314 Dresden, Germany}
\author{J. Wosnitza}
\affiliation{Helmholtz-Zentrum
Dresden-Rossendorf, Bautzner Landstr. 400, D-01314 Dresden, Germany}
\affiliation{Institut f\"ur Festk\"orper- und 
Materialphysik und W\"urzburg-Dresden 
Cluster of Excellence ct.qmat, TU Dresden, 
01062 Dresden, Germany}

\date{\today}

\begin{abstract}
We report Alfv\'en-wave experiments with liquid 
rubidium at the Dresden High Magnetic Field Laboratory (HLD). 
Reaching up to 63\,T, the pulsed magnetic field exceeds the 
critical value of 54\,T
at which the Alfv\'en speed becomes equal to the sound speed
(plasma-$\beta$ unity).
At this threshold  we observe a period doubling of an applied 
8\,kHz CW excitation, a clear footprint for a parametric 
resonance between magnetosonic waves and Alfv\'en waves.

\end{abstract}

\maketitle

Since their discovery by Hannes Alfv\'en in 1942 \cite{Alfven1942}, 
Alfv\'en waves have played an ever increasing role in understanding 
astrophysical and fusion-related plasmas. They 
are, in particular, one of the main ingredients 
to explain the dramatic heating of the solar 
corona \cite{Tomczyk2007} and to 
accelerate the solar wind 
\cite{Kasper2019}, they are found 
in the Earth's ionosphere \cite{Lysak2013}, and they 
are being employed for the heating of fusion plasmas 
\cite{Fasoli1995,Intrator1995}. More generally, Alfv\'en waves serve as a 
reference paradigm for a variety of  waves and 
instabilities in rotating magnetized plasmas or liquid metals, 
in particular the  
magnetorotational instability \cite{Balbus1991,Stefani2006} 
and torsional oscillations in the Earth's outer core 
\cite{Gillet2010}.

Laboratory experiments on Alfv\'en waves in 
liquid mercury \cite{Lundquist1949} and sodium \cite{Lehnert1954}
had started 
soon after Alfv\'en's theoretical prediction.  
Since those early times, many Alfv\'en wave experiments 
have been carried out, both with liquid metals 
\cite{Jameson1964,Druyvesteyn1968,Iwai2003} and, 
more extensively, 
with plasmas \cite{Gekelman1999,Amatucci2006}. Recently, 
liquid-metal experiments were resumed in order 
to study Alfv\'en waves with pulsed excitations 
\cite{Alboussiere2011}, as well as torsional 
Alfv\'en waves in spherical geometry \cite{Tigrine2019}.

\begin{figure}
\includegraphics[width=\columnwidth]{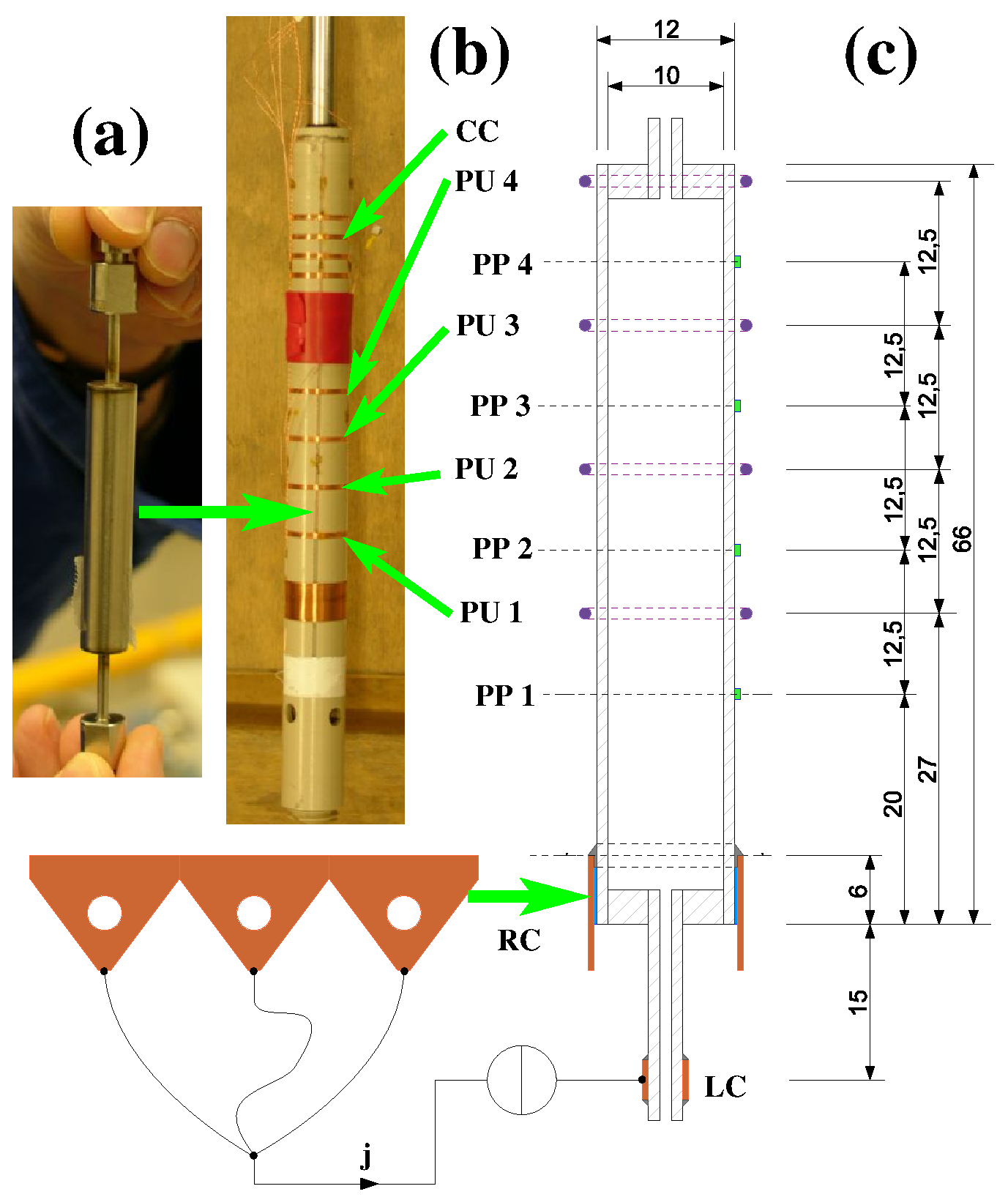}
\caption{Experimental setting: (a) Stainless-steel container 
filled with rubidium. (b) Holder with 
four pick-up coils (PU 1 - PU 4), 
and four compensation coils (CC). 
(c) Geometrical details of the construction. 
PP 1 - PP 4 denote four electric potential probes 
at the container. The three 
orange triangles indicate the rim contacts (RC) encircling 
the bottom part of the container. LC is the lower 
contact. All sizes are in mm.
}
\end{figure}

The recent progress in generating pulsed magnetic fields 
beyond  $B=90$\,T \cite{Wosnitza2007,Zherlitsyn2012} 
opens up a
completely new prospect for Alfv\'en-wave experiments with 
liquid metals. It is at those fields 
(51\,T for caesium and 
54\,T for rubidium) that 
the Alfv\'en speed $v_{\rm a}=B/(\mu_0 \rho)^{0.5}$ in higher 
alkali metals crosses 
the sound speed $c_{\rm s}$. 
This threshold, sometimes called 
plasma-$\beta$ unity (in plasma 
physics, $\beta$ is defined as the ratio of thermal to magnetic pressure), 
is of key importance 
for the mutual transformability of Alfv\'en waves
and (slow and fast) magnetosonic waves 
\cite{Zaqarashvili2006a,Warmuth2005}, and 
its crossing at the so-called ``magnetic canopy'' 
has been made responsible for the heating of the corona 
\cite{Hollweg1980,Bogdan2003}. One particular effect 
predicted to occur at this
threshold is the transfer of the energy of magnetosonic waves 
of a given frequency into Alfv\'en waves of half that frequency
(period doubling) that was explained in  
terms of parametric resonance, or swing 
excitation \cite{Zaqarashvili2006}.

\begin{figure}
\includegraphics[width=0.97\columnwidth]{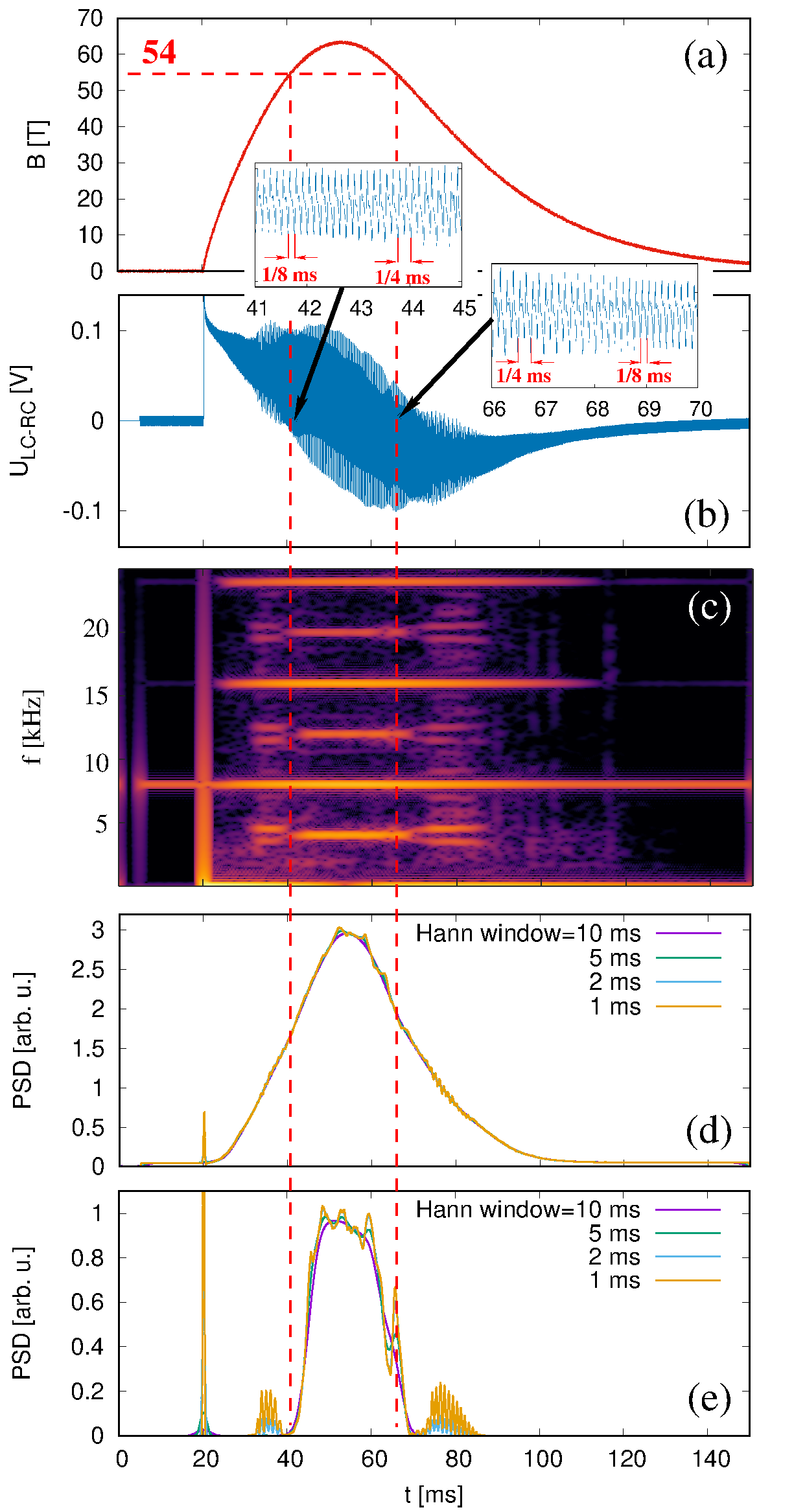}
\caption{Time dependence of the pulsed magnetic field (a), and 
 of the voltage measurements at the lower contact (b). The 
 red dashed lines indicate the instants where the critical 
 field strength of 54\;T 
is crossed. The insets of (b) detail the two transition regions, 
where
the double-period signal starts and ceases to exist. Gabor 
transform (c) of the signal from (b), with a Hann window 
width of 5\,ms. Amplitude of the 8\,kHz (d) and the 4\,kHz (e) 
stripe from (c), for different choices of the 
von Hann window width. The PSD units 
of (d) and (e) are arbitrary, but consistent
among each other.}
\end{figure}

\begin{figure}
\includegraphics[width=0.97\columnwidth]{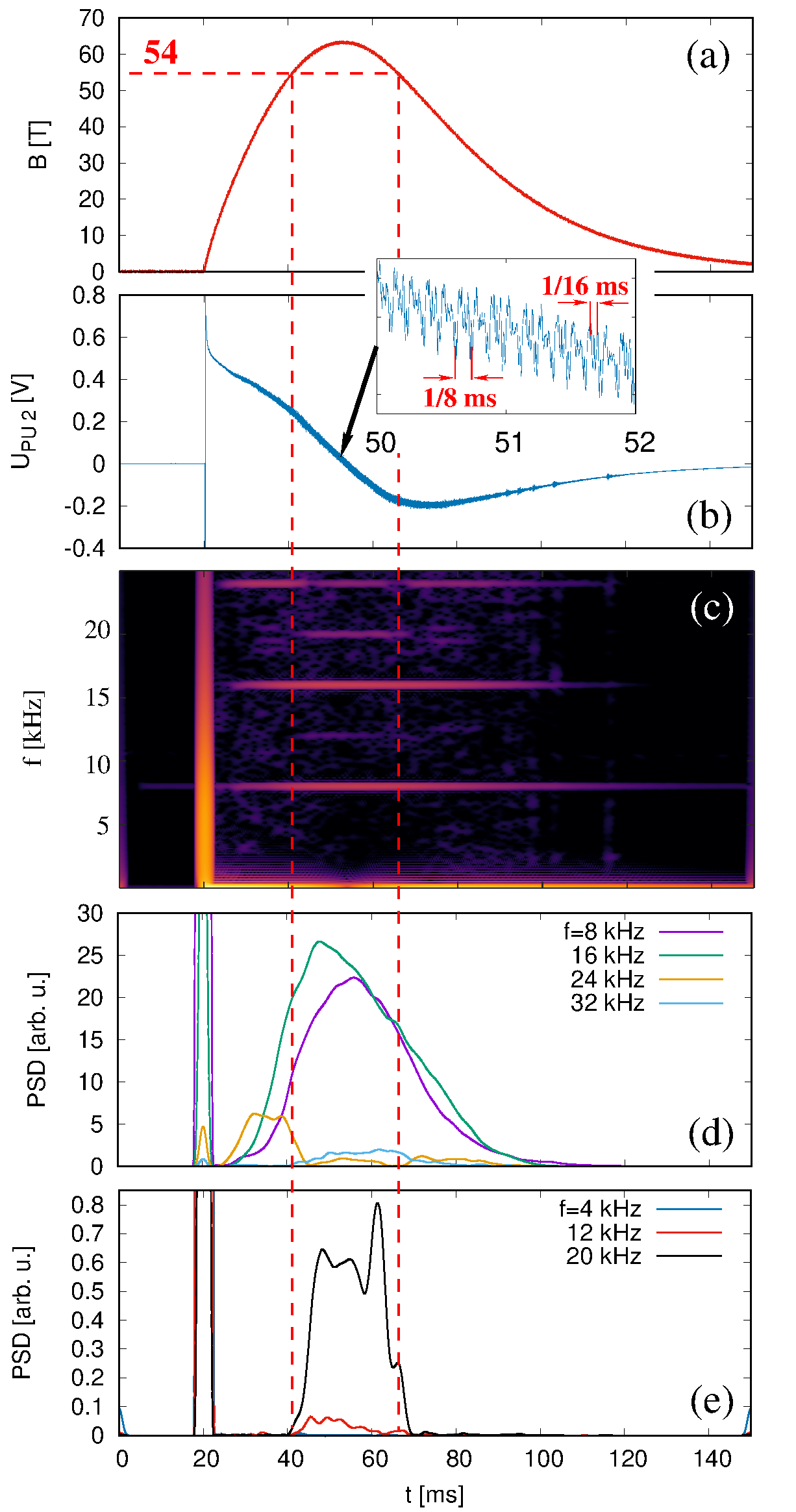}
\caption{Time dependence of the applied magnetic field (a), and 
of the (compensated) voltage measurement at the 
pick-up coil PU 2 (b). The 
red dashed lines indicate the instants where the critical field 
strength of 54\,T 
is crossed. The inset of (b) shows the region of highest field, 
where a significant contribution of a 16\,kHz signal is visible. 
Gabor transform (c) of the signal from (b), with a von Hann window 
width of 5\,ms. Amplitudes of the 
8\,kHz, 16\,kHz, 24\,kHz, and 32\,kHz  
stripes (d) and of the 4\,kHz, 12\,kHz, and 20\,kHz 
stripes (e) from (c). The PSD units 
of (d) and (e) are arbitrary, but consistent
among each other.}
\end{figure}

In this Letter, we report experimental evidence for the
occurrence of this period-doubling effect in an 
Alfv\'en-wave experiment with  liquid 
rubidium. 
Using one of the pulsed-field coils at 
the Dresden High Magnetic 
Field Laboratory (HLD), the achieved maximum field value 
of 63\,T 
exceeds the threshold $v_{\rm a}=c_{\rm s}$, reached at 54\,T, 
by 17 per cent. 
Yet, the pulsed character of the field entails 
various technical challenges, among them a huge 
contribution of the magnetic-flux derivative to the voltage 
induced in the pick-up coils (an effect that is, by and large,  
overcome by using compensation coils), 
and a significant pressure buildup in the liquid metal 
leading to an uncontrolled background flow field. 
A further challenge is the limited size of the available 
experimental volume: 
the cold-bore diameter in the 70\,T long pulse coil of the HLD 
\cite{Wosnitza2007} is 24\,mm.  Since the 
copper-alloy coil of the large 
magnet is immersed into liquid nitrogen, the (warm) rubidium 
experiment has to be shielded thermally from this cold surrounding 
by a double-wall Dewar lance which reduces the available 
diameter further. A holder for the pick-up coils is also necessary, 
so that the outer diameter of the stainless steel container for the 
Rubidium is 12 mm. As the pressure in the liquid, arising 
from the steep increase of the pulsed magnetic field, 
reaches values in the order of 
50\,bar, a container wall thickness of 1 mm is necessary, which 
reduces the ultimately available diameter of the 
rubidium column to 10 mm. 
The height of this column, limited by the homogeneity 
region of the magnetic field, has been chosen as 60 mm. The 
central part of the set-up, with the container, the 
coil holder, the pick-up coils and the electric-potential
probes is visualized in Fig 1.

Figure 2 shows some chief results of one experimental 
run, carried out at a temperature of $50^\circ\text{C}$ 
at which rubidium is liquid.
After releasing, at $t=20$\,ms, the energy 
from the capacitor bank (charged with
22\,kV) the axial magnetic field [Fig. 2(a)] 
increases swiftly to attain its maximum value of 63.3\;T
at $t=53$\,ms. From there on, the field declines
slowly, reaching a value of 2.1\,T at the end of the 
interval considered here ($t=150$\,ms).
The period during which the critical value of 54\,T is 
exceeded ranges 
from 40.5\,ms until 66\,ms, as indicated by the 
dashed red lines. 
During the entire experiment, a sinusoidal 
CW current (not shown) 
with constant amplitude of 5\,A and frequency 
8\,kHz was applied
between the lower contact (LC) and the three 
contacts (RC) encircling the lower rim 
of the container
[illustrated at the lower left part of Fig. 1(c)]. 
The corresponding current density ${\bf j}_{\rm r}$, which 
is concentrated in the bottom layer of the rubidium where
it is basically directed in radial direction, generates 
together with the  strong vertical field ${\bf B}_{\rm z}$ 
an azimuthal Lorentz force density 
${\bf f}_{\rm \varphi}={\bf j}_{\rm r} \times {\bf B}_{\rm z}$  
that is supposed to drive a torsional Alfv\'en 
wave in the fluid.

The voltage $U_{\rm{LC-RC}}$ 
measured between the contacts LC and RC  
comprises three contributions: First, a  
significant electro-motive force (emf) 
$\bf{v}_{\varphi} \times {\bf B}_{\rm z}$ arising 
from the interaction of the toroidal 
velocity ${\bf v}_{\varphi}$ 
of the generated torsional (Alfv\'en) wave with the 
axial magnetic field ${\bf B}_{\rm z}$.  
Second, the usual Ohmic voltage drop 
[giving an amplitude of approximately 
5\,mV, as seen at at the start and end of the magnetic-field pulse
when 
the magnetic field is close to zero, Fig. 2(a)]. Third,  
a certain long-term trend resulting from 
the time derivative of the pulsed field
(which is still induced in the
wire system used for the electric contacts that is 
oriented not perfectly radially).

Already without having performed detailed 
numerical analyses, we can at least 
plausibilize the measured 
oscillation amplitude of this emf: 
with an applied current 
amplitude of 5\,A, we obtain (for the given 
contact geometry) a current density of 
$j_{\rm r} \approx 100$\,kA/m$^2$, leading 
(with $B_{\rm z}=50$\,T and a density of 
$\rho=1.49$\,kg/m$^3$) 
to an azimuthal acceleration of 
$a_{\varphi} \approx 3000$\,m/s$^2$. 
Acting over half an excitation period  (1/16\,ms), 
this acceleration 
can generate flow velocities of $v_{\varphi}\approx 20$\,cm/s, 
which in turn induce (integrated over the container's 
radius of $r=5$\,mm) 
an emf of $v_{\varphi} B_{\rm z} r \approx 50$\,mV. This 
estimation is 
well compatible with the amplitude of the voltage oscillation  
as seen in the high-field segment of Fig. 2(b).

The measured voltage is now analyzed by means of 
a windowed Fourier transform (or Gabor transform, using the
``ltfat'' toolbox in Octave \cite{ltfat}). For a 
von Hann window 
width of 5\,ms, Fig. 2(c) shows the arising spectrogram 
(i.e., the Power Spectral Density (PSD) 
over time), restricted here to 
the frequency range $0-25$\,kHz, 
with a resolution of 100\,Hz. The dominant 
feature of this spectrogram is, not surprisingly, 
the 8\,kHz signal, whose
time dependence is separately plotted in Fig. 2(d) 
for four different widths of the von Hann window. 
Since 
the amplitude of the velocity induced 
emf increases smoothly with $B_{\rm z}$, there is 
a negligible 
dependence on the window width.
Another unsurprising feature seen in Fig. 2(c) is 
the appearance of various overtones of 8\,kHz.

What is surprising, however, is the appearance of a 
strong {\it period-doubling}
4\,kHz line (and its overtones) which shows up nearly 
exclusively in the time interval when the field 
is equal to, or larger than, the critical value of 
54\,T. 
Figure 2(e) illustrates this restriction quite clearly.
Notable is a slight asymmetry in time: 
after crossing the threshold, the 4\,kHz mode seems to 
require some time to fully develop, while it extends 
slightly beyond the end of the interval. 
Note also, exclusively at the (slowly) 
decreasing branch, the special peak appearing 
exactly at $B=54$\,T, which is not visible 
on the (steeply) increasing branch.
The remarkable cleanliness and the significant PSD amplitude
of this 4\,kHz mode (reaching one third of the PSD of the 
8\,kHz mode), is also highlighted in the two insets 
of Fig. 2(b) where the initial onset and the later 
decay of that mode are shown.
Without an explanation 
at hand, we also hint to the obvious splitting of this 4\,kHz mode into 
two sidebands, which occurs shortly before and after 
the critical threshold [Fig. 2(c)], in a 
field interval between 30 and 54\,T.

With Fig. 3, we turn now to the voltage measurements 
from the pick-up coil PU 2 that is located slightly 
above mid-height 
of the container (the signals of the other PU's are similar). 
Despite a significant reduction due to the use 
of a compensation coil, the dominant part of this signal 
[Fig. 3(b)]
is still coming from the time derivative of the pulsed magnetic 
field [note the similarity of this shape with the 
shape of the low-frequency part in Fig. 2(b)].
Superposed on that, we observe a high-frequency
part comprising the usual 8\,kHz, and a 
second-harmonic 16\,kHz signal 
of approximately the same amplitude. How to explain that mixture?
To start with, the voltage in the pick-up coils results from 
time derivatives of azimuthal currents ${\bf j}_{\varphi}$
which are typically produced by the induction effect 
${\bf v}_{\rm r} \times {\bf B}_{\rm z}$
of radial flow components ${\bf v}_{\rm r}$ with the 
axial field ${\bf B}_{\rm z}$. That flow component, in turn, is 
intimately connected with radial and axial gradients of the
pressure $p$, and is, therefore, indicative of the presence of 
(fast and slow) magnetosonic waves.
Hence, the appearance of the 16\,kHz signal is a well-known 
consequence of the
quadratic dependence of the pressure on the 
magnetic-field perturbation of the Alfv\'en wave, as 
already discussed for the 
gallium experiment of Iwai et al. \cite{Iwai2003}. 
As also observed in \cite{Iwai2003}, the original
driving frequency (8\,kHz in our case) is still present,
and both frequencies have rather comparable 
amplitudes, as evidenced in Fig. 3(d). The third and fourth 
harmonics, however, are already much smaller.

Remarkably here is the nearly complete absence
of any 4\,kHz signal, and only very minor
contributions from the higher $(2n+1)\times 4$\,kHz harmonics
[note the significant differences of the PSD units between
Figs. 3(d) and 3(e)]. This absence of any 
{\it pressure related} 4\,kHz signal
is in stark contrast to the
quite significant share of this frequency band  
in the voltage measurements at $B>54$\,T [Fig. 2(e)], 
which we attribute to the emergence of a new torsional 
Alfv\'en wave at the 54\,T threshold related to 
$v_{\rm a}=c_{\rm s}$.

\begin{figure}
\includegraphics[width=\columnwidth]{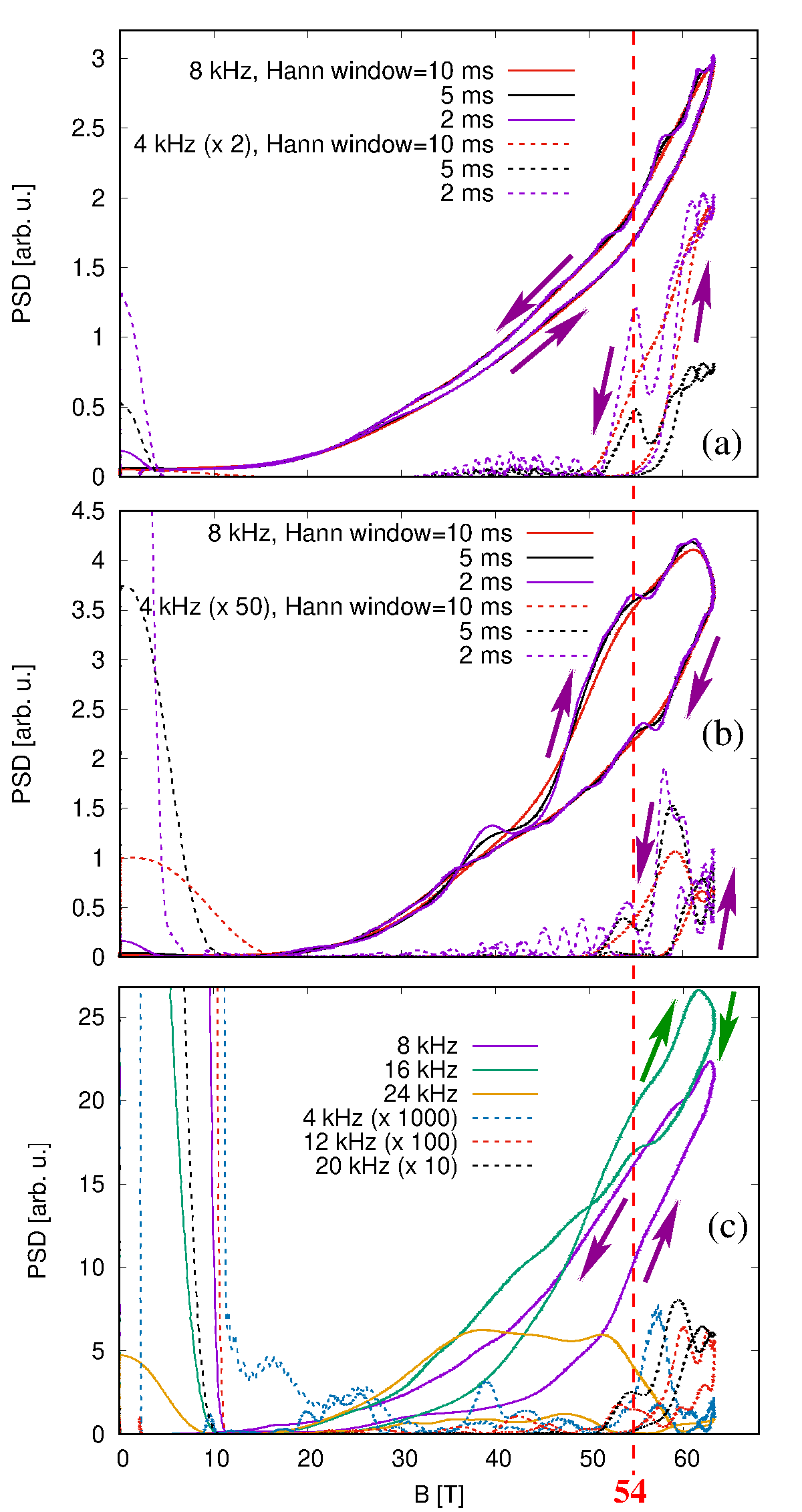}
\caption{Dependence of  various measured signals on the 
magnetic field. (a) PSD of the 8\,kHz and the 4\,kHz 
coefficients of the voltage between the
contacts from Figs. 2(d) and 2(e)m for three 
different Hann windows. 
(b) The same for the voltage measured between PP 2 and PP 3. 
(c) PSD of the 8, 16, and 24\,kHz signals from Fig. 3(d), and
of the 4, 12, and 20\,kHz signals from Fig. 3(e). The arrows 
indicate rising and falling flanks of the pulsed 
magnetic field. }
\end{figure}

In Fig. 4, we translate those time series
into corresponding magnetic-field dependencies, 
including both the 
increasing and decreasing branches. 
Based on the data from 
Fig. 2, Fig. 4(a) 
confirms, first, the expected quadratic dependence of 
the PSD of the driven 8\,kHz mode (with only a slight 
difference between the increasing and the decreasing 
branch) and, second,  a clear peak of the 
4\,kHz signal at 54\,T coming from the decreasing branch,
and another peak behind, coming from the increasing branch.
Without proof, we assume that this asymmetry 
is due to the transient character of the experiment.

With Fig. 4b we add a further result from an electric
potential measurement between the contacts 
PP 2 and PP 3. 
Basically, the underlying signal (not shown here) 
and the Gabor
transform look quite similar to those presented 
in Fig. 2, except
that the typical voltage is smaller by a factor 100
(this is due to the fact that the emf is mainly produced 
between the axis and the rim of the container, 
while PP 3 - PP 2  measures only some
residual voltage difference in $z$ direction). 
We observe a similar behaviour as discussed
before, with a mainly 
quadratic dependence of the 8\,kHz signal on $B$ (with some stronger
up-down asymmetry, however), and a steep increase of the 4\,kHz 
signal at and beyond 54\,T.

Figure 4(c) shows the PSD dependence on the magnetic field for the 
signal of the pick-up coil PU 2 from Fig. 3. Evidently the
8 and 16\,kHz signals look quite similar, with some differences 
in the up-down relation which  still remain to be understood. 
At any rate, the 4, 12, and 20\,kHz curves are extremely 
small (note, in particular, the factor 1000
by which which the 4\,kHz signal must be multiplied to become visible).

To the best of our knowledge, we have carried out the 
first systematic Alfv\'en wave experiment where the ``magical" 
threshold plasma-$\beta$ unity has 
been crossed.
In contrast to the many plasma experiments, which are usually 
carried out at small values of $\beta$
(but see \cite{Cekic1992,Okada2001,Flanagan2020} for exceptions), 
liquid metal experiments
were up to present limited to the 
region with $\beta \gg 1$, or $v_{\rm a} \ll c_{\rm s}$. 
Using liquid rubidium in the high 
pulsed fields available at the HLD, we were 
able to exceed this threshold by some 17 percent. 
Our main observation was the 
sudden appearance (at $v_{\rm a} = c_{\rm s}$) 
of a period-doubled 4\,kHz mode, which we  
interpret as a new torsional Alfv\'en wave 
arising by parametric resonance, or swing excitation 
\cite{Zaqarashvili2006}, from the 
usual 8\,kHz magnetosonic wave.
Admittedly, the first realization 
of this astrophysically important phenomenon  
under strongly transient experimental conditions
demands for substantial numerical support.   
Complementary experiments with
longer pulses would also be desirable
for a better understanding of the 
observed time asymmetries.

 
We acknowledge support of the HLD at HZDR, 
a member of EMFL, and the DFG through the 
W\"urzburg-Dresden Cluster of Excellence on 
Complexity and Topology in Quantum Matter - 
ct.qmat (EXC 2147, Project No. 390858490).
F.S. acknowledges further support 
by the European Research Council (ERC)
under the European Union's Horizon 2020 Research and Innovation
Programme (Grant No. 787544). We thank J\"urgen H\"uller 
for his assistance in the adventurous filling procedure 
of the rubidium container, and Frank Arnold, 
Carsten Putzke, Karsten Schulz and Marc Uhlarz 
for their help in preparing  and carrying out the experiment.

\end{document}